\documentclass[onecolumn,aps,preprint,showpacs,amsmath,amssymb]{revtex4}
\usepackage{graphicx}
\usepackage{latexsym}
\usepackage{dcolumn} 
\usepackage{stmaryrd} 
 
\usepackage{bm}

\begin{document}
\newcommand{\eq}{\begin{equation}}                                                                         
\newcommand{\eqe}{\end{equation}}             

\title{Self-similar solutions of the three dimensional Navier-Stokes equation} 

\author{ I. F. Barna}
\address{$^a$ KFKI Atomic Energy Research Institute of the Hungarian Academy 
of Sciences, Thermohydraulics Department\\ (KFKI-AEKI), H-1525 Budapest, P.O. Box 49, Hungary} 

\date{\today}

\begin{abstract} 

In this article we will present pure three dimensional analytic  solutions for the Navier-Stokes and the continuity equations in Cartesian coordinates. The key idea is the three-dimensional generalization of the well-known self-similar Ansatz of Barenblatt.
A geometrical interpretation of the Ansatz is given also. 
The results are the Kummer functions or strongly related. Our final formula is compared with other results obtained 
from group theoretical approaches. 
 
\end{abstract}

\pacs{47.10.ad}
\maketitle

To describe the dynamics of viscous incompressible fluids the Navier-Stokes (NS)  partial differential equation (PDE) 
together with the continuity equation have to be investigated.
In Cartesian coordinates and Eulerian description  these equations 
have the following form:  
\begin{eqnarray}
{\bf{\nabla}} {\bf{v}} = 0,  \nonumber  \\               
 {\bf{v}}_t  + ({\bf{v}}\nabla){\bf{v}} = \nu \triangle 
{\bf{v}} - \frac{\nabla p}{\rho} + a               
\label{nav} 
\end{eqnarray}

where ${\bf{v}}, \rho, p, \nu, a $ denote respectively the three-dimensional velocity field, density, pressure, kinematic viscosity 
and an external force (like gravitation) of the investigated fluid.
(To avoid further misunderstanding we use $a$ for external field instead of the letter g which is reserved for a self-similar solution.)  
In the following $ \nu, a $ are parameters of the flow. 
For a better transparency in the following we use the coordinate notation ${\bf{v}}(x,y,z,t) = u(x,y,z,t),v(x,y,z,t),w(x,y,z,t)$ and for the scalar  pressure variable $p(x,y,z,t)$ 
\begin{eqnarray}
u_x + v_y + w_z = 0 \nonumber \\ 
u_t + uu_x + vu_y + wu_z = \nu(u_{xx}+u_{yy}+u_{zz}) -
\frac{p_x}{\rho}    \nonumber \\ 
 v_t + uv_x + vv_y + wv_z = \nu(v_{xx}+v_{yy}+v_{zz}) -
\frac{p_y}{\rho}   \nonumber \\ 
w_t + uw_x + vw_y + ww_z = \nu(w_{xx}+w_{yy}+w_{zz}) 
-\frac{p_z}{\rho} + a. 
\label{nav2}
\end{eqnarray}
The subscripts mean partial derivations.  
According to our best knowledge there are no analytic solution  
for the most general three dimensional case. 
However, there are various examination techniques available in the literature.  
Manwai  \cite{manwai} studied the N-dimensional $(N \ge 1)$ radial 
Navier-Stokes equation with different kind of viscosity and pressure 
dependences and presented analytical blow up solutions.	
His works are still 1+1 dimensional (one spatial and one time dimension)
investigations.  Another well established and popular investigation method is based on 
Lie algebra there are numerous studies available.  
Some of them are even for the three dimensional case,  
for more see \cite{lie}. Unfortunately, no explicit solutions are shown and 
analyzed there. Fushchich {\it{et al.}} \cite{fus} construct a complete set of ${\tilde{G}}(1,3)$-inequivalent Ans\"atze of codimension 1 for the NS system, they present 19 different analytical solutions for one or two space dimensions. 
They last solution is very closed to our one but not identical, we will come back to these results later. 
Further two and three dimensional studies based on group analytical method were presented by Grassi \cite{grassi}. 
They also present solutions which look almost the same as ours, but they consider only 2 space dimensions.  
We will compare these results to our one at the end of the paper.  

 Recently, Hu {\it{et al.}} \cite{hu} presents a study where 
symmetry reductions and exact solutions of the (2+1)-dimensional NS were presented. 
Aristov and Polyanin \cite{arist} use various methods  like Crocco transformation,  generalized separation of variables or the method of functional separation of variables for the NS and present large number of new classes of exact solutions. 
Sedov in his classical work \cite{sedov}* presented analytic solutions for the tree dimensional spherical NS equation 
where all three velocity components and the pressure have polar angle dependence ($\theta$) only. Even this kind of restricted symmetry led to a non-linear coupled ordinary differential equation system with has a very rich mathematical structure.    

Beyond the NS system there are other important and popular PDEs which attract much interest and investigation. 
The applied methods are the same there, too. Without completeness we mention some examples. 
For one dimensional cubic-quintic nonlinear Schr\"odinger equation a quite general self-similar type of solution
$\psi(z,t)= u(z,t)exp[iv(z,t)]$ was applied where u and v are real functions \cite{wu}. The results are  
analytic solutions for an external potential with variable coefficients. 
A more general type of this Ansatz $u(z,t) = A(z)U[T(z,t)exp(i \varphi(z,t))]$ was used 
with success to get chirped and chirp-free self-similar cnoidal solitary wave solutions \cite{dai} for the same equation.  
Such solutions can be generalized for multi dimensional spatial coordinates. There are analytic solitary wave solutions available
for the (3+1) dimensional   Gross-Pitaevskii equation with the following Ansatz $\psi = u(x,y,z,t)R(t)exp[i b(t)(x^2+y^2+z^2) ]$  \cite{gao}.

From basic textbooks the form of the one-dimensional self-similar Ansatz is well-known \cite{sedov,barenb,zeld} 
\eq 
T(x,t)=t^{-\alpha}f\left(\frac{x}{t^\beta}\right):=t^{-\alpha}f(\eta), 
\label{self}
\eqe 
where $T(x,t)$ can be an arbitrary variable of a PDE and $t$ means time and $x$ means spatial 
dependence.
The similarity exponents $\alpha$ and $\beta$ are of primary physical importance since $\alpha$  represents the rate of decay of the magnitude $T(x,t)$, while $\beta$  is the rate of spread 
(or contraction if  $\beta<0$ ) of the space distribution as time goes on.
The most powerful result of this Ansatz is the fundamental or 
Gaussian solution of the Fourier heat conduction equation (or for Fick's
diffusion equation) with $\alpha =\beta = 1/2$. These solutions are visualized on figure 1. for time-points $t_1<t_2$. 
In the pioneering work of Leray  \cite{leray} in 1934 at the end of the manuscript he asks whether it is possible to construct 
self-similar solutions to the NS system in ${\bf{R}}^3$  in the form of $p(x,t) = \frac{1}{T-1}P(x/\sqrt{T-t})$ and 
${\bf{v}}(x,t) = \frac{1}{\sqrt{T-t}}{\bf{V}}(x/{\sqrt{T-t}})$. 
In 2001 Miller {\it{et al.}} \cite{miller} proof the nonexistence of singular pseudo-self-similar solutions of the NS system
with such kind of solutions. 
Unfortunately, there is no direct analytic calculation with the 3 dimensional self-similar generalization of this Ansatz in the literature.  We will show later on that in our case the time dependence has the same exponents as showed above.  
 
Applicability of this Ansatz is quite wide and comes up in various 
transport systems \cite{sedov,barenb,zeld,kers,barn,barna2}. 
\begin{figure} 
\scalebox{0.4}{
\rotatebox{0}{\includegraphics{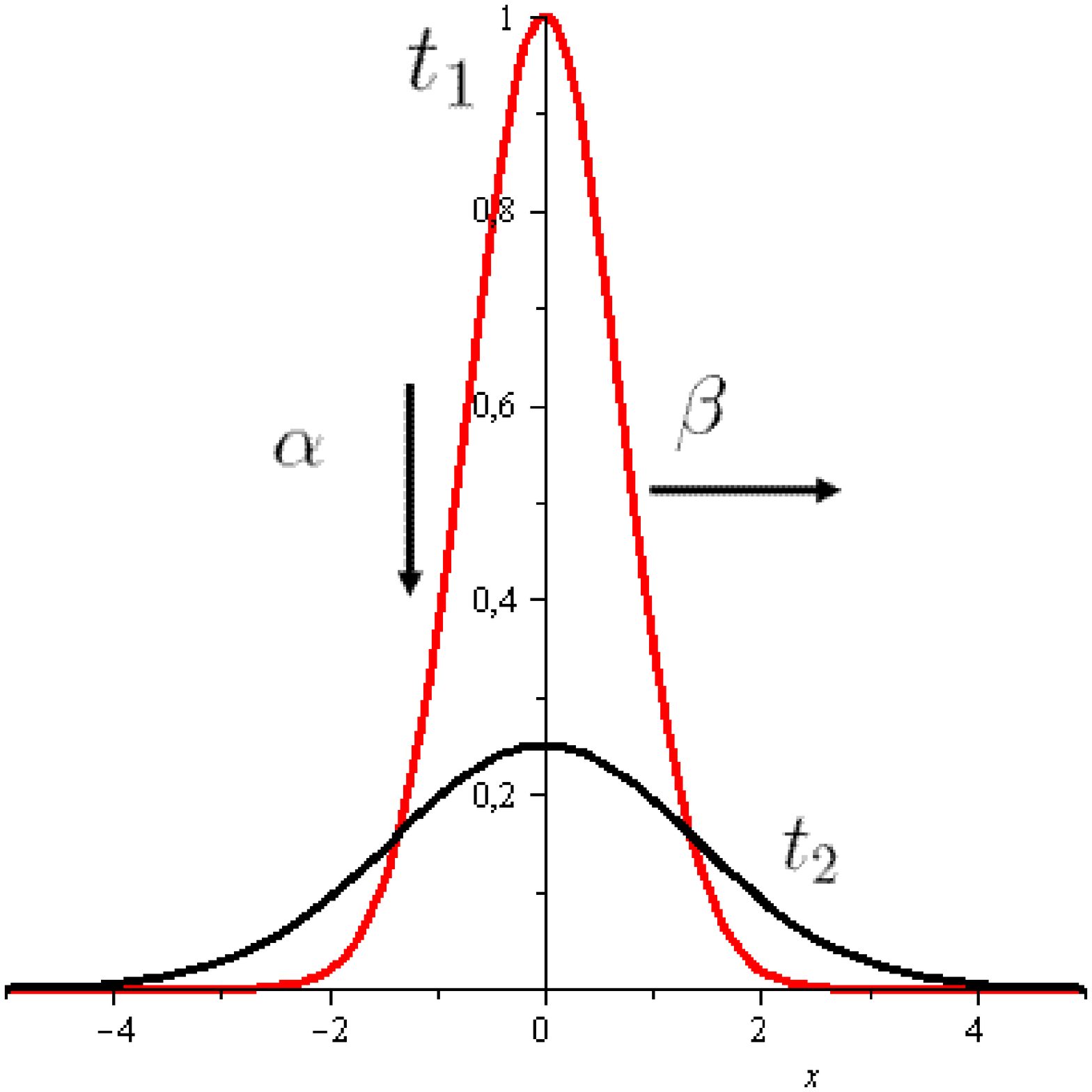}}}
\caption{A self-similar solution of Eq. (\ref{self}) for $t_1<t_2$.
The presented curves are Gaussians for regular heat conduction.}	
\label{egyes}       
\end{figure}
This Ansatz can be generalized for two or three dimensions in various ways
one is the following 
\eq
u(x,y,z,t) = t^{-\alpha}f\left(\frac{F(x,y,z)}{t^\beta}\right) := t^{-\alpha}f\left(\frac{x+y+z}{t^\beta}\right):=t^{-\alpha}f(\omega)
\eqe
where $F(x,y,z)$ can be understood as an implicit parameterization of a two dimensional surface. 
If the function $F(x,y,z)=x+y+z=0$ which is presented on figure 2. then it is an implicit form of a plane in three dimensions. At this point we can give a geometrical interpretation of the Ansatz. Note that the dimension of $F(x,y,z)$ still have to be 
a spatial coordinate.  With this Ansatz we consider all the x coordinate of the velocity field ${\bf{v}_x} = u $ where the sum of the spatial coordinates are on a plane on the same footing. We are not considering all the $R^3$ velocity field but a plane of the ${\bf{v}}_x$ coordinates as an independent variable. 
The Navier-Stokes equation - which is responsible for the dynamics - maps this kind of velocities which are on a surface to another geometry.
In this sense we can investigate the dynamical properties of the NS equation  
truly.  
\begin{figure}  
\scalebox{0.35}{
\rotatebox{0}{\includegraphics{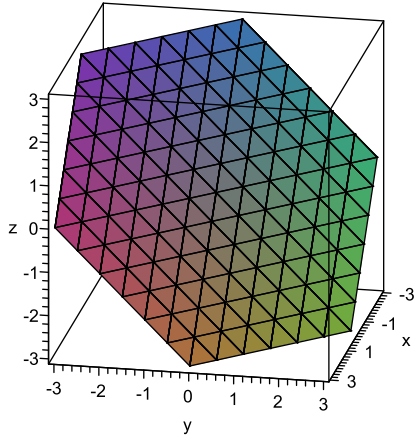}}}
\caption{The graph of the $x+y+z=0$ plane.}	
\label{kettes}       
\end{figure}
In principle there are more possible generalization of the Ansatz 
available. One is the following: 
\eq
u(x,y,z,t) = t^{-\alpha}f\left(\frac{\sqrt{x^2+y^2+z^2-a}}{t^\beta}\right):=
t^{-\alpha}f(\omega)
\eqe
which can be interpreted as an Euclidean vector norm or $L^2$ norm. 
Now we contract all the x coordinate of the velocity field $u$ (which are on a surface of a sphere with radius a)  to a simple spatial coordinate. 
Unfortunately, if we consider the first and second spatial derivatives 
and plug them into the Navier-Stokes equation we cannot get a pure
$\eta $ dependent ordinary differential equation(ODE) system some explicit $x,y,z$ or $t$ dependence tenaciously remain. 
For a telegraph-type heat conduction equation both these Ansatzes are useful to get solutions for the two dimensional case \cite{barna2}.  
 
Now we concentrate on the first Ansatz (4) and search the solution of the Navier-Stokes PDE system in the following form:  
\begin{eqnarray}
u(x,y,z,t) = t^{-\alpha} f\left(\frac{x+y+z}{t^{\beta}}\right), \hspace*{3mm} 
v(x,y,z,t) = t^{-\gamma} g\left(\frac{x+y+z}{t^{\delta}}\right), 
\nonumber \\
w(x,y,z,t) = t^{-\epsilon} h\left(\frac{x+y+z}{t^{\zeta}}\right), \hspace*{3mm} 
p(x,y,z,t) = t^{-\eta} l\left(\frac{x+y+z}{t^{\theta}}\right).
\label{ans}
\end{eqnarray}
Where all the exponents $\alpha,\beta,\gamma,\delta, \epsilon,
\zeta, \eta,\theta$ are real numbers. (Solutions with integer exponents are called self-similar solutions of the first kind, non-integer exponents 
mean self-similar solutions of the second kind.) 
The functions $f,g,h,l$ are arbitrary and will be evaluated later on.  
According to Eq. (\ref{nav2}) we need to calculate all the first time derivatives of the velocity field, all the first and second spatial 
derivatives of the velocity fields and the first spatial derivatives of the 
pressure. All these derivatives are not presented in details. 
Note that both Eq. (\ref{nav2}) and Eq. (\ref{ansatz}) have 
a large degree of exchange symmetry in the coordinates $x,y$ and $z$.   
Later we want to have an ODE system 
for all the four functions $f(\omega), g(\omega), h(\omega), l(\omega)$ 
which all have to have the same argument $\omega $. This dictates the 
constraint that $\beta = \delta = \zeta = \theta$ have to be the same
real number which reduces the number of free parameters, (let's use the $\beta$ from now on $\omega = \frac{x+y+z}{t^{\beta}})$. From this constrain follows that e.q.  $ u_x =\frac{f'(\omega)}{t^{\alpha+\beta}}  \approx  v_y =\frac{f'(\omega)}{t^{\gamma+\beta}} $  where prime means derivation with 
respect to $\omega$.   This example shows the hidden symmetry of this construction which may helps us. 
For the better understanding we present the second equation of (\ref{nav2})
after the substitution of  the Ansatz (\ref{ans}). 
\begin{eqnarray}
 -\alpha t^{-\alpha-1} f(\omega) - \beta t^{-\alpha-1}f'(\omega)\omega  + t^{-2\alpha - \beta}f(\omega)f'(\omega) + t^{-\gamma-\alpha-\beta}g(\omega)f'(\omega) + \nonumber \\ t^{-\epsilon-\alpha-\beta}h(\omega)f'(\omega) = \nu 3 t^{-\alpha-2\beta} f''(\omega) - \frac{t^{-\mu-\beta}l'(\omega)}{\rho}.
\label{ode1}
\end{eqnarray} 

To have an ODE which only depends on $\omega$ (which is now the new variable instead of time t and the radial components) all the time dependences 
e.g. $t^{-\alpha-1}$ have to be zero OR all the exponents have to be the same.  
After some algebra it comes out that all the six exponents $\alpha-\zeta$ included for the velocity filed (the first three functions in Eq. (\ref{ansatz})) have to be $+1/2$. The only exception is the term with 
the gradient of the pressure. There $\eta =1$ and $\theta =1/2$ have to be. 
Now in Eq. (\ref{ode1}) each term is multiplied by $t^{-3/2}$. 
Self-similar exponents with the value of $+1/2$ are well-known from  
the regular Fourier heat conduction (or for the Fick's diffusion) equation and gives back the fundamental solution which is the usual Gaussian function. 
For pressure the $\eta =1$ exponent means, a two times quicker decay rate 
of the magnitude than for the velocity field.

\begin{figure} 
\scalebox{0.32}{
\rotatebox{0}{\includegraphics{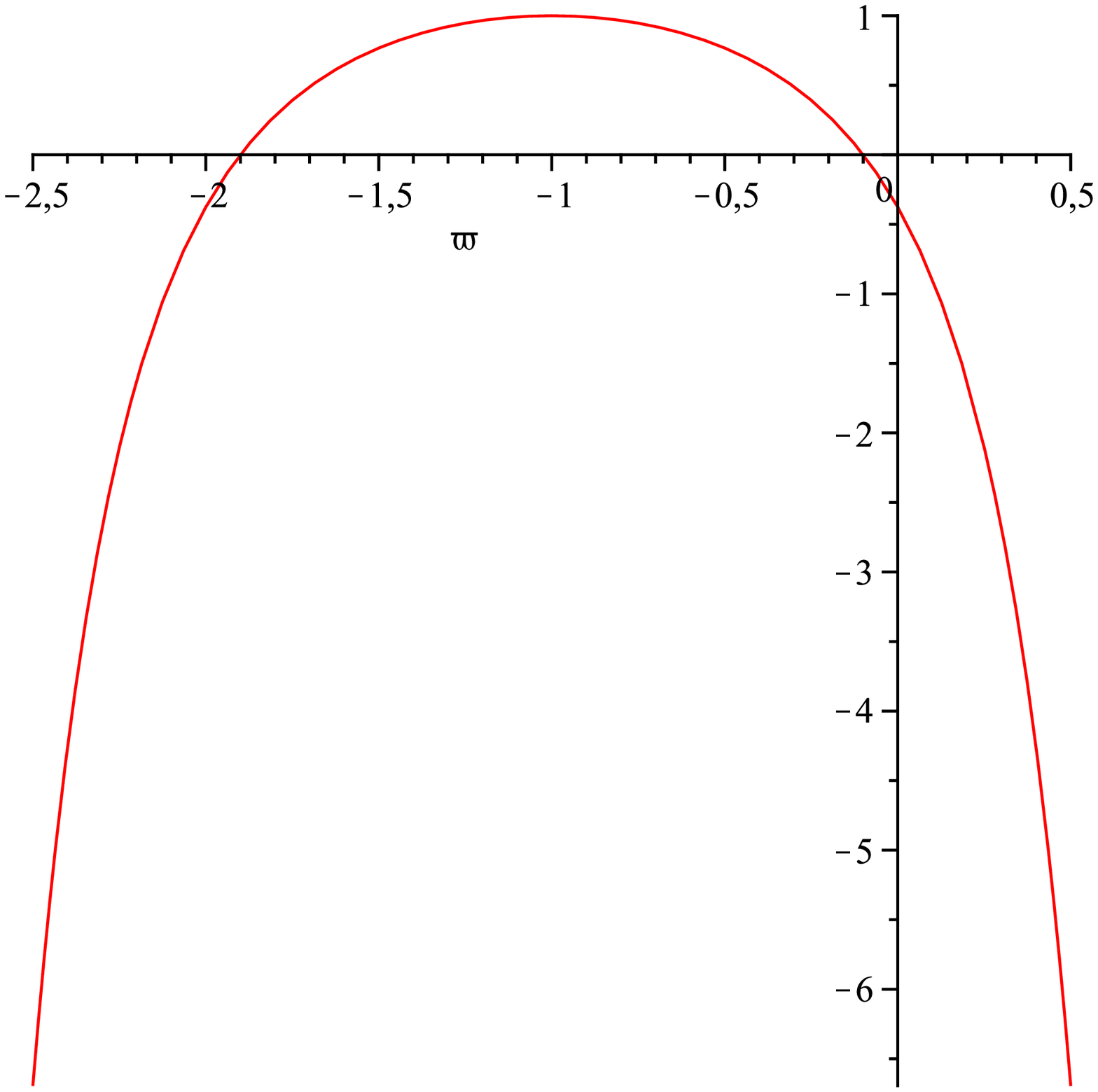}}}
\caption{The $KummerM \left( -\frac{1}{4},\frac{1}{2},\frac{(\omega+c)^2}{6\nu} \right)$ function for 
$c = 1$, and $\nu = 0.1$.  }	
\label{harmas}       
\end{figure}

\begin{figure} 
\scalebox{0.32}{
\rotatebox{0}{\includegraphics{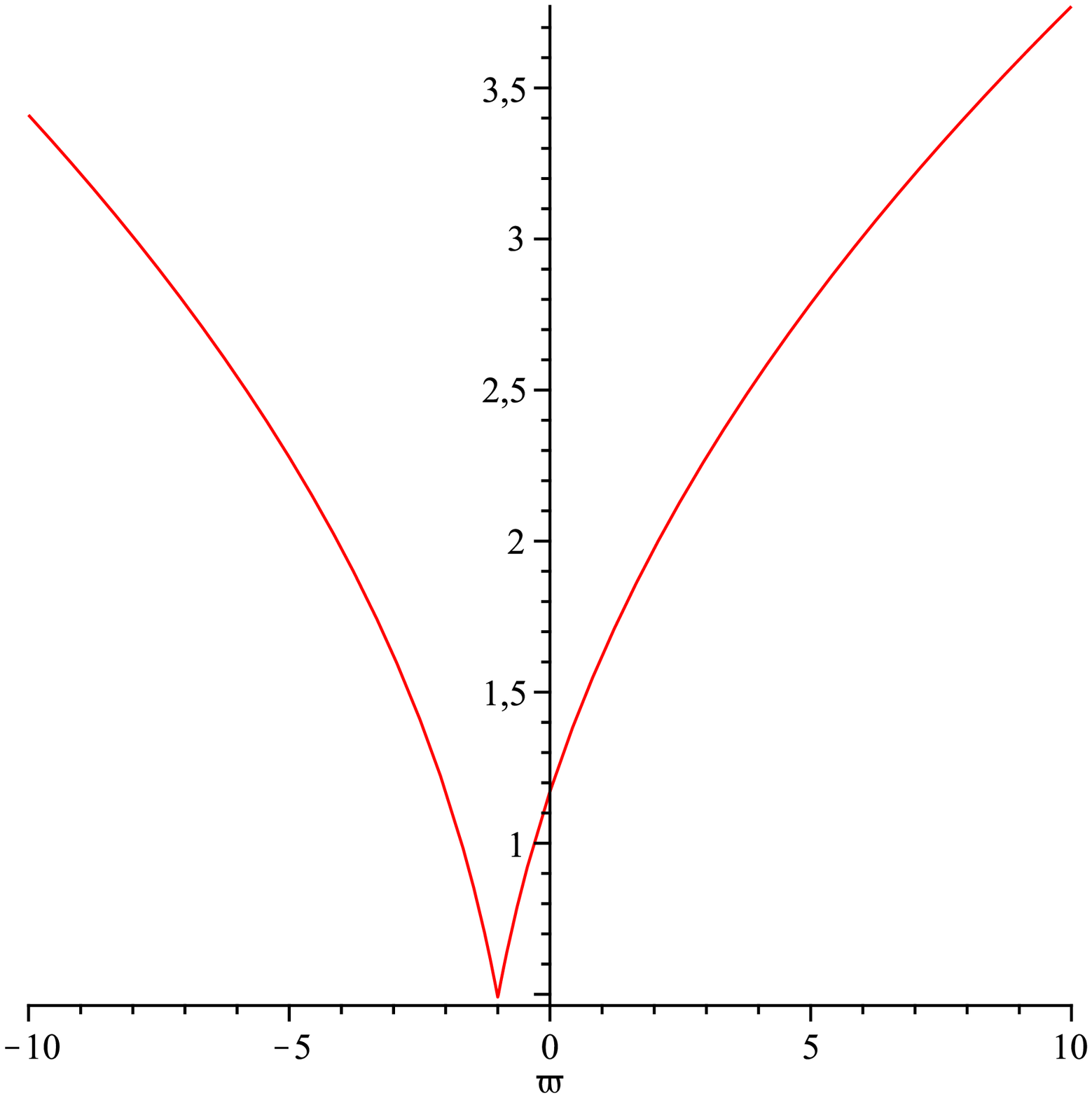}}}
\caption{The $KummerU \left( -\frac{1}{4},\frac{1}{2},\frac{(\omega+c)^2}{6\nu} \right)$ function for 
$c = 1$, and $\nu = 0.1$.   }	
\label{negyes}       
\end{figure}
\begin{figure} 
\scalebox{0.32}{ 
\rotatebox{0}{\includegraphics{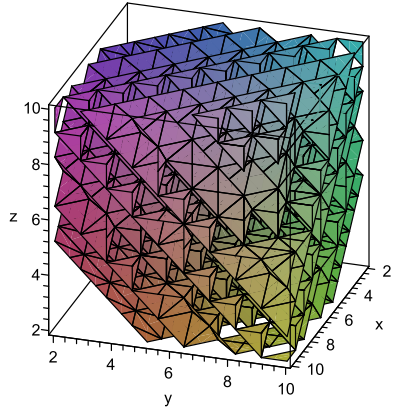}}}
\caption{The implicit plot of the self-similar solution Eq. (\ref{vx}).
Only the KummerU  function is presented for $t=1, c_1=1, c_2 = 0, a=0,  
c = 1$, and $\nu = 0.1$.}	
\label{negyes}       
\end{figure}

Now we may write down the concrete form of the Ansatz (\ref{ans}) 
\begin{eqnarray}
u(x,y,z,t) = t^{-1/2} f\left(\frac{x+y+z}{t^{1/2}}\right) = 
t^{-1/2}f(\omega), \hspace*{3mm} 
v(x,y,z,t) = t^{-1/2} g(\omega), 
\nonumber \\
w(x,y,z,t) = t^{-1/2} h(\omega), \hspace*{3mm} 
p(x,y,z,t) = t^{-1} l(\omega), 
\label{ansatz}
\end{eqnarray}

and the corresponding coupled ODE system

\begin{eqnarray}
f'(\omega) + g'(\omega) + h'(\omega)& = &0 \nonumber \\ 
 -\frac{1}{2}f(\omega) - \frac{1}{2}\omega f'(\omega)  + 
[f(\omega) + g(\omega) + h(\omega)]f'(\omega) &=& 3\nu f''(\omega)  - \frac{l'(\omega)}{\rho} \nonumber \\
-\frac{1}{2}g(\omega) - \frac{1}{2}\omega g'(\omega)  + 
[f(\omega) + g(\omega) + h(\omega)]g'(\omega) &=& 3\nu g''(\omega)  - \frac{l'(\omega)}{\rho} \nonumber \\
-\frac{1}{2}h(\omega) - \frac{1}{2}\omega h'(\omega)  + 
[f(\omega) + g(\omega)+ h(\omega)]h'(\omega) &=& 3\nu h''(\omega)  - \frac{l'(\omega)}{\rho} +a. 
\label{ode1}
\end{eqnarray}

From the first (continuity) equation we automatically get 
\eq
f(\omega) + g(\omega) + h(\omega) = c, \hspace*{3mm}  and \hspace*{3mm}
f''(\omega) + g''(\omega) + h''(\omega) =0 
\eqe
where c is proportional with the constant mass flow rate. Implicitly, larger c means 
larger velocities. 
From the second equation we can express $-\frac{l'}{\rho}$ and can 
substitute it into the third and fourth equation. 
After some algebra we arrive at 
\begin{eqnarray} 
\frac{f(\omega)-g(\omega)}{2} +\frac{\omega(f'(\omega)-g'(\omega))}{2} + 3\nu(f''(\omega)-g''(\omega)) \nonumber \\
+[f(\omega)+g(\omega)+h(\omega)](g'(\omega)-f'(\omega))  = 0 \nonumber \\
\frac{f(\omega)-h(\omega)}{2} +\frac{\omega(f'(\omega)-h'(\omega))}{2} + 3\nu(f''(\omega)-h''(\omega)) \nonumber \\ 
+[f(\omega)+g(\omega)+h(\omega)](h'(\omega)-f'(\omega)) + a = 0. \nonumber \\
	\label{ketto}  
\end{eqnarray}
Now inserting $f''(\omega)=-g''(\omega)-h''(\omega)$, $f'(\omega)=-g'(\omega)-h'(\omega)$ and $f(\omega) = c-g(\omega)-h(\omega)$ 
we get the final equation 
\eq
9\nu f''(\omega) - 3(\omega +c)f'(\omega) + \frac{3}{2}f(\omega) - \frac{c}{2} +a = 0. 
\label{kummerdif} 
\eqe
The solutions are the Kummer functions \cite{abr}  
\eq
f(\omega) = c_1 \cdot KummerU \left(-\frac{1}{4},\frac{1}{2},\frac{(\omega+c)^2}{6 \nu} \right)
+ c_2 \cdot KummerM \left( -\frac{1}{4},\frac{1}{2},\frac{(\omega+c)^2}{6 
\nu} \right) + \frac{c}{3} -\frac{2a}{3}
\label{kum}
\eqe 
where $c_1$ and $c_2$ are integration constants. 
The KummerM function is defined by the following series 
\eq
M(a,b,z) = 1+ \frac{az}{b} + \frac{(a)_2 z^2}{(b)_2 2!} + ... +
\frac{(a)_n z^n}{(b)_n n!}
\eqe
where $(a)_n$ is the Pochhammer symbol 
\eq
(a)_n = a(a+1)(a+2)...(a+n-1), (a)_0 =1. 
 \eqe
The KummerU function is defined from the KummerM function 
via the following form
\eq
U(a,b,z) = \frac{\pi}{sin(\pi b)} 
\left[ \frac{M(a,b,z)}{\Gamma(1+a-b)\Gamma(b)} 
- z^{1-b}\frac{M(1+a-b,2-b,z)}{\Gamma(a)\Gamma(2-b)} \right]   
\eqe
where $\Gamma()$ is the Gamma function.
Exhausted mathematical properties of the Kummer function can be found 
in \cite{abr}.

Note, that the solution depends only on two parameters where the $
\nu$ is the viscosity, and $c$ is proportional with the mass flow rate. 
Figure 3 and figure 4 show the KummerM and KummerU function for $c=1$ and $\nu = 0.1$, respectively. 
For stability analysis we note that the power series which is applied to calculate the Kummer function has a 
pure convergence and a  30 digit accuracy was needed to 
plot the KummerU function, otherwise spurious oscillations occurred on the 
figure.  Note, that for $\omega =6.5$ the KummerM goes to infinity, and 
$\omega \rightarrow \infty$  KummerU function goes to $\infty$ 
which is physically hard to understand, which means that the velocity field goes to infinity as well. 
  
The complete self-similar solution of the x coordinate of the velocity  
reads 
\begin{eqnarray}
u(x,y,z,t) =  &t&^{-1/2}f(\omega) =  t^{-1/2}\left[c_1 \cdot KummerU\left(\frac{-1}{4},\frac{1}{2}, 
\frac{((x+y+z)/t^{1/2} + c)^2}{6\nu} \right) \right] + \nonumber \\ 
 &t&^{-1/2}\left[ c_2 \cdot KummerM \left( -\frac{1}{4},\frac{1}{2},\frac{((x+y+z)/t^{1/2}    +c)^2}{6 
\nu} \right) + \frac{c}{3} -\frac{2a}{3}  \right].
\label{vx}
\end{eqnarray}
On figure 5 an implicit plot of Eq. (\ref{vx}) is visualized. 
The KummerU function was presented only, the used parameters are
the following $c_1= 1,c_2=0,t=1,c=1,\nu=0.1, a=0$. Note, that the initial flat surface of figure 2 is mapped into a complicated topological surface via the NS dynamical equation. The following phenomena happened, 
an implicit function is presented, we already mentioned that all the $x+y+z =0 $ points considered to be the same. 
Therefore we got a multi-valued surface because for a fixed x numerical value various y+z combinations give the same 
argument inside the Kummer function.  Unfortunately, this effect  is hard to visualize. 
This can be understand as a kind of fingerprint of a turbulence-like phenomena which is still remained in the equation. 
An initial simple single-valued plane surface is mapped into a very complicated multivalued surface. Note, that for a larger value (now we presented KummerU() = 2 case)  or for larger flow rate (c=1) the surface got even more structure. Therefore figure 5 presents only a principle.  
At this point we can also give statements about the stability of this solution, the solution the Kummer functions are fine, but 
for larger flow values a more precise and precise calculation of the solution surface is needed which means 
larger computational  effort which is well known from the application of the NS equation. 

From the integrated continuity equation $(f=c-g-f)$ we automatically get an implicit formula for the other two velocity components
\begin{eqnarray}
 v(x,y,z,t) + &w&(x,y,z,t) = - t^{-1/2}\left[c_1 \cdot KummerU\left(\frac{-1}{4},\frac{1}{2}, 
\frac{((x+y+z)/t^{1/2} + c)^2}{6\nu} \right) \right] - \nonumber \\ 
 &t&^{-1/2}\left[ c_2 \cdot KummerM \left( -\frac{1}{4},\frac{1}{2},\frac{((x+y+z)/t^{1/2}    +c)^2}{6  
\nu} \right) + \frac{c}{3} -\frac{2a}{3}  \right] + c.
\end{eqnarray} 
For explicit formulas of the remaining two velocity components  the 
two ODEs of (\ref{ketto}) have to be integrated. 
For $v(x,y,z,t) = t^{-1/2}g(\omega) $ the ODE is the following
\eq
 -3\nu g''(\omega) + g'(\omega)\left(-\frac{\omega}{2}+c\right) -\frac{g(\omega)}{2} + F(f''(\omega),f'(\omega),f(\omega)) = 0 
\label{hard}
\eqe
where $F(f''(\omega),f'(\omega),f(\omega)) $ contains the combination of 
the first and second derivatives of the Kummer functions. 
This is a second order linear ODE and the solution can be obtained with the following general quadrature
\eq 
g(\omega) = \left[c_2 + \int \left\{ \frac{-c_1 + \int F(f''( \omega),f'(\omega),f(\omega)) d\omega \cdot  exp\left(\frac{-\omega^2/4+c\omega}{-3\nu} \right) }{3\nu}    \right\}   d\omega \right] 
exp\left(\frac{-\omega^2/4+c\omega}{3\nu} \right). 
\eqe 
For the sake of simplicity we present the formulas of the first and second derivatives of the KummerU functions only   
\eq
\frac{d}{d\omega} KummerU(a,b,\omega)  =  
\frac{(\omega+a-b)KummerU(a,b,\omega) -KummerU(a-1,b,\omega) }{\omega} 
\eqe
and 
\begin{eqnarray}
\frac{d^2}{d\omega^2}KummerU(a,b,\omega) = 
\frac{1}{\omega^2} \left[ a \left\{ \omega a U(a+1,b,\omega) 
- \omega U(a+1,b,\omega)b + \omega U(a+1,b,\omega) \right. \right.
  \nonumber \\ 
\left. \left.  -a U(1+a,b,\omega)b +
U(a,b,\omega)b + U(a+1,b,\omega)b^2  - U(a+1,b,\omega)b  \right\}  \right]. 
\end{eqnarray}
   Unfortunately, we could not find any closed form for $v(x,y,z,t)$ 
and for $w(x,y,z,t)$. Only $v$ the $x$ coordinate of the velocity ${\bf{v}}$ field can be evaluated in a closed form.   

As we mentioned at the beginning there are analytic solutions available in the literature which are very similar to our one. 
Fushchich {\it{et al.}} \cite{fus} present 19 different solutions for the full three dimensional NS and continuity equation.
(For a better understanding we used the same notation here as well.) 
For the last (19th) solution they apply the following Ansatz of 
\eq
u(z,t)= \frac{f(\omega)}{\sqrt{t}}, \hspace*{2mm} v(y,z) =  \frac{g(\omega)}{\sqrt{t}} + \frac{y}{t}, \hspace*{2mm}
 w(z,t) = \frac{h(\omega)}{\sqrt{t}}, \hspace*{2mm} p(t,z) =  \frac{ l(\omega)}{\sqrt{t}}  
\eqe
where $\omega = z/{\sqrt{t}}$ is the invariant variable. The obtained ODE is very similar to ours  (\ref{ode1})  
\begin{eqnarray} 
h'(\omega) +1 &=& 0 \nonumber \\ 
-\frac{1}{2}(f(\omega) + \omega f'(\omega)) + h(\omega) f'(\omega) &=& f''(\omega), \nonumber \\ 
\frac{1}{2}(g(\omega) + \omega g'(\omega)) + h(\omega) g'(\omega) &=& g''(\omega), \nonumber \\ 
-\frac{1}{2}(h(\omega) + \omega h'(\omega)) + h(\omega) h'(\omega) +l'(\omega) &=& f''(\omega).  
\end{eqnarray} 
The solutions are 
\begin{eqnarray}
f(\omega) &=& (\frac{3}{2}\omega -c)^{-1/2}exp \left [ -\frac{1}{6} ( \frac{3}{2}\omega-c)^2 \right]  w \left[ -\frac{1}{12},\frac{1}{4},\frac{1}{3}(\frac{3}{2}\omega -c)^2 \right]  \nonumber \\ 
g(\omega) &= &(\frac{3}{2}\omega -c)^{-1/2}exp \left [ -\frac{1}{6} ( \frac{3}{2}\omega-c)^2 \right]  w \left[ -\frac{5}{12},\frac{1}{4},\frac{1}{3}(\frac{3}{2}\omega -c)^2 \right]  \nonumber \\ 
h(\omega) &=& -\omega + c \nonumber \\ 
l(\omega) &=& \frac{3}{2}c \omega - \omega^2 + c_1
\end{eqnarray}
where $w$ is the Whitakker function, c and $c_1$ are integrational constants. 
Note that the Whitakker and the Kummer functions are strongly related  to each other \cite{abr}*  
\eq
w(\kappa,\mu,z) = e^{-1/2z}z^{1/2+\mu} KummerM(1/2+\mu-\kappa, 1+2\mu,z).  
\eqe
More details can be found in the original work \cite{fus}.   

As a second comparison we show the results of \cite{grassi}.  
They also have a modified form of (\ref{nav2}) which is the following 
\begin{eqnarray}
U_{1t} + cU_1 + U_2U_{1y} + U_3U_{1z} - \nu(U_{1yy}+U_{1zz})    &=& 0, \nonumber \\  
U_{2t} + U_2 U_{2y} + U_3U_{2z}  +\pi_y- \nu ( U_{2yy}+U_{2zz})    &=& 0, \nonumber \\ 
U_{3t} + U_2U_{3y} + U_3U_{3z} +\pi_z- \nu(U_{3yy}+U_{3zz})    &=& 0, \nonumber \\
U_{2y} + U_{3z} + c & =& 0  
\end{eqnarray}
where $U_i, i=1..3$ are the velocity components 
$U_i(y,z,t)$ and $\pi$ is the pressure, c stands for constants, $\nu$ is viscosity and additional 
subscripts mean derivations.  
After some transformation they get a linear PDA as follows
\eq
U_{1t} + k_1y U_{1y} + (\sigma-k_1z)U_{1z} - \nu(U_{1yy}+U_{1zz})    = 0 
\eqe
it is convenient to look the solution in the form of 
\eq
U_1 = Y(y)T(z)\Phi(t). 
\eqe
Note, that they also consider the full 3 dimensional problem, but the velocity filed has a restricted 
two dimensional(y,z) coordinate dependence. 
There are additional conditions but the general solution can be presented
\begin{eqnarray}
\Phi &=& c_1 exp(c_2)t  \nonumber \\
Y &=& c_3 M \left(  -c_4, \frac{1}{2}, \frac{y^2}{2\nu}\right) + 
yc_5 M \left( \frac{1}{2}-c_4, \frac{3}{2}, \frac{y^2}{2\nu}\right) \nonumber \\ 
T &\approx& M \left( c_6,\frac{1}{2},\frac{z^2}{2\nu} \right) + zM \left( \frac{1}{2}-c_6,\frac{3}{2},
\frac{z^2}{2\nu}\right)
\label{olasz}
 \end{eqnarray}
where M is the Kummer M function as was presented below. 
The exact solution in \cite{grassi} (4.10a-4.10c)* contains more constants as presented here. 
It is not our goal to reproduce the full calculation of \cite{grassi} (which is not our work)
we just want to give a guideline to their solution vigorously emphasising  that our solution is very similar to the presented one.
Note that in both results the arguments of the Kummer M function (\ref{kum}) and (\ref{olasz}) are proportional to the square of the radial component divided by the viscosity, additionally one of the parameters is 1/2.   
As a last word we just would like to say, (as this example clearly shows) that the Lie algebra method is not the 
exhaustive method to find all the possible solutions of a PDA.  \\

{\it{In summary:}} We introduced and gave a geometrical interpretation of a 
three-dimensional self-similar Ansatz. We applied it to the three-dimensional 
Navier-Stokes equation in Cartesian coordinates. The question of another Ans\"atze was mentioned briefly as well. Some part of the results could be written as Kummer functions. 
Unfortunately, some other parts of the results could not be written in closed forms.  Further work is in progress, (we still have some hope) to learn something new from 
Eq. (\ref{hard}).  
We compared our results with other analytic solutions obtained from various Lie algebra studies. 
The structure of the result - the implicit coordinate dependence of the Kummer function - was analyzed as well.  
We hope that even this moderate result can give any simulating impetus to 
the investigation of the Navier-Stokes equation.  
Our solution can have some real interest and can be used as a test case 
for various numerical methods or commercial computer packages like Fluent or CFX. \\ 
The paper is dedicated to my first mathematics teacher "Sanyi B\'acsi".



\end{document}